\documentclass[twocolumn,showpacs,aps]{revtex4}

\usepackage{graphicx} 
\usepackage{dcolumn}  
\usepackage{bm}       

\input psfig.sty 

\input rotate
\input epsf

\def\bea{\begin{eqnarray}}
\def\eea{\end{eqnarray}}
\def\ben{\begin{equation}}
\def\een{\end{equation}}
\def\benu{\begin{enumerate}}
\def\enu{\end{enumerate}}

\def\n{n}

\def\sss{\scriptscriptstyle\rm}

\def\g{_\gamma}

\def\l{^\lambda}

\def\1var{(\bx_1...\bx\N)}

\def\half{\frac{1}{2}}

\def\br{{\bf r}}

\def\bx{{x}}

\def\x{_{\sss X}}
\def\c{_{\sss C}}

\def\xc{_{\sss XC}}

\def\N{_{\sss N}}

\def\pot{^{\rm pot}}

\def\pot{^{\rm pot}}

\def\unif{^{\rm unif}}

\def\up{_\uparrow}
\def\dn{_\downarrow}

\def\KS{Kohn-Sham~}
\def\DFT{density functional theory~}
\def\dft{density functional theory~}

\def\sph_int{ {\int d^3 r}}

\def\dnarrow{\downarrow} 
\def\up{_\uparrow  } 
\def\dn{_\downarrow} 

\def\n{n}

\def\pot{v}

\def\d3r{ d^3 r \; }

\def\sclup{\alpha}
\def\scldn{\beta}

\def\lamup{\lambda\up}
\def\lamup{\lambda\up}

\def\ups{_{\uparrow   \sclup } } 
\def\dns{_{\downarrow \scldn } }

\def\xcup{ _{ {\sss XC} \uparrow}       }

\def\xcdn{ _{ {\sss XC} \downarrow}       }

\def\xclup{_{ {\sss XC} }({\lambda\up}) }

\def\xlup{_{ {\sss X} }  ({\lambda\up}) }

\def\spiel{local spin density approximation (solid line), 
generalized gradient approximation (PBE, dashed line),
BLYP (bars)}

\def\spielplus{\spiel, self-interaction corrected LSD 
(short dashes)}

\begin{document} 
\headheight 50pt

\preprint{RUTGERS DFT GROUP: pre-print MWB02}

\title{Scaling the spin densities separately in
density functional theory} 

\author{R.J. Magyar}
\affiliation{Department of Physics, Rutgers University,
136 Frelinghuysen Road, Piscataway, NJ 08854}

\author{T.K. Whittingham}
\affiliation{Department of Chemistry, Rutgers
University, 610 Taylor Road, Piscataway, NJ 08854} 

\author{K.  Burke}
\affiliation{Department of Chemistry and Chemical Biology, 
Rutgers University, 610 Taylor Road, Piscataway, NJ 08854} 


\begin{abstract} 
Coordinate scaling of each spin
density separately is considered in spin density
functional theory.  A virial
theorem relates the spin-scaled correlation energy to the
spin-scaled correlation potentials.  
An
adiabatic connection formula expresses energies
at different spin interaction strengths in terms of spin
scaling.  
Several
popular approximate functionals are evaluated on
the spin-scaled densities of atoms and of the
uniform electron gas.   
The differences between this and uniform scaling are discussed.
\end{abstract}


\date{\today}
\maketitle


\section{Introduction} \label{s:intro} 

Density functional theory combines accuracy and 
celerity in a computational scheme 
which has long been used by
solid state physicists and has become
popular in quantum chemistry \cite{Kb99}.  
The Hohenberg-Kohn
theorem \cite{HK64} demonstrates that the electron 
density uniquely characterizes a ground
state electronic structure problem so 
that the total energy is a functional of the
density.  
This idea suggests the computationally expedient  
Kohn-Sham scheme \cite{KS65}
where only one part of the total energy
functional must be approximated. This is the 
exchange-correlation energy, $E\xc$, and the calculational 
accuracy
of \DFT is limited by the accuracy of approximations to
$E\xc$; therefore, 
improvements in this functional are of great import.

Exact constraints limit the possible forms of
approximations to $E\xc$ and provide guidance
for the construction of approximations \cite{PBE96}.
For total electron density
functionals, Levy and Perdew \cite{LP85} discovered a
set of important scaling and integral requirements that
the exact functional must satisfy.
They introduced the concept of uniform coordinate density scaling.
This scaling takes a density $\n(\br)$ into
\bea
\n\g (\br) &=& \gamma^3
\n(\gamma\br),~~~~0 \leq \gamma < \infty,
\label{scaling_law}
\eea
and is a natural way to explore the behavior of density
functionals. Many properties of the exact functional
have been found by studying its behavior under this
scaling \cite{L91,ZLP93,LGb95}.
For example, the exchange energy
changes as
\ben
E\x [\n_\gamma] = \gamma E\x[\n].
\label{scaling_ex}
\een
All commonly-used approximations such as the local 
density approximation (LDA), the generalized 
gradient approximation (GGA), and hybrids of GGA with
exact exchange \cite{B93} 
satisfy this relation, some by 
construction.
But for correlation, only inequalities can be derived,
\ben
E\c [\n\g] > \gamma E\c[\n]~~~~~~~~~(\gamma > 1),
\label{ec_scaling}
\een
which some approximations satisfy \cite{PW92,PBE96} while
others do not \cite{LYP88}.
Levy and Perdew further showed that translational invariance implies
virial theorems relating potentials to energies.  An example is
\ben
E\x[\n] =
 -\int\d3r\n(\br) \; \br\cdot\nabla
\pot\x[\n] (\br)
\label{ex_virial}
\een
with a more complicated corresponding relation for correlation.
The virials have been used to construct energy densities
directly from potentials \cite{CLB98}.
Exact statements about
exact functionals are nontrivial and extremely
useful in the construction and analysis of
approximate functionals.
Functionals that violate these exact conditions
are unlikely to give reliable and physical results
when applied to wide ranges of materials.

On the other hand, modern density functional calculations
do not employ {\em density} functionals but rather
use {\em spin}-density functionals. 
The basic idea is to
replace $E\xc[\n]$ with $E\xc[\n\up,\n\dn]$ so that the
universal functional depends explicitly on both the up and down
spin electron densities.  Formal justification for this scheme 
was first given by von Barth and Hedin \cite{BH72}  and later by 
Rajagopal and Calloway \cite{RC73} although some fundamental 
questions remain \cite{CV0}.  
There are several compelling reasons for using spin density 
functionals instead of total
electron density functionals.  
Spin density functionals can more accurately 
describe systems with odd numbers of electrons \cite{JGP93}.  
They also allow the treatment of electrons in
collinear magnetic fields and 
yield magnetic response properties \cite{MM0}.
Accurate calculation of these properties would be far more difficult
in total-density functional theory, because
a local
spin density functional is a non-local total density
functional.  
One cannot exaggerate how useful spin \dft has been in 
accurately and efficiently calculating physical properties. 

To help develop improved spin-density functionals,
it would be of great interest 
to develop a formalism that probes
the spin dependence of functionals and yields
exact conditions about their spin-dependence.
Here, we investigate whether scaling 
techniques developed
for total density functionals can be generalized to spin 
density functional theory.
Towards this end, we generalize 
uniform coordinate scaling of the density, Eq. (\ref{scaling_law}) , to
separate scaling of spin-densities:
\bea
\n_{\uparrow\sclup}(\br)
&=&
\sclup^3 \n\up
(\sclup\br),~~~~0 \leq \sclup < \infty \nonumber\\
\n_{\downarrow\scldn}(\br)
&=&
\scldn^3 \n\dn
(\scldn\br),~~~~0 \leq \scldn < \infty. 
\label{ss_law}
\eea  
According  to this  scheme,  a  spin-unpolarized system  becomes
spin-polarized for $\sclup\neq \scldn$.
There  are many other  ways  we could  have chosen to alter the spin
densities.
For  example,  we  could
require that the total  density remain constant while the polarization
changes.  However, such 
a  transformation would require introduction of ensembles because fractions
of electrons would be changing  spin. 
The present scheme is simply the logical extension of coordinate scaling
to separate spin densities. 
  
A principal result of this work is that, under such a transformation, 
a spin dependent 
virial theorem holds true.  
This theorem can be used to carefully 
check the convergence of spin-density functional calculations. 
It also provides a method for calculating the exact dependence
of the correlation energy on spin-density for model
systems for which accurate Kohn-Sham potentials have been found.
Finally, it can be integrated over the scale factor 
$\sclup$ to give a new formal expression 
for the functional, $E\xc$.

Considerable progress has been made in \dft by writing 
$E\xc$ as an integral over a coupling constant $\lambda$
in what is called the adiabatic connection relationship \cite{LP75,GL76}.
For example, the success of hybrid functionals 
such as B3LYP \cite{B93,Bb93} 
can be understood in terms of this adiabatic connection \cite{BEP97,PEB96}.
The adiabatic connection is simply related to uniform
coordinate scaling \cite{Wb97,B97}.
By analogy,
we relate spin scaling to a spin-coupling constant
integration, and we define a suitable generalization 
for this definition with a coupling constant for each spin density.

We illustrate our formal results with several cases.
For the uniform electron gas, we can perform this scaling
essentially exactly.
We show how this transformation relates energies 
to changes in spin-polarization.  
In this case, considerable care must be taken to deal with 
the extended nature of the system.
We also show the results of spin scaling 
densities of small atoms using presently popular approximations.
We close with a discussion of the fundamental difficulty 
underlying this spin scaling approach.

Throughout, we use atomic units
($e^2=\hbar=m_e=1$), so that all energies are in
Hartrees and all lengths in Bohr radii.
We demonstrate all scaling relationships by 
scaling the up spin densities.  Results for 
scaling the down spin are obtained in a similar fashion.  
Just change the spin label!


\section{Separate spin scaling theory}
\label{s:theory}

The first interesting property of the spin scaling 
transformation, Eq. (\ref{ss_law}), is that it conserves the total number 
of electrons globally even though the scaled spin 
density might tend towards zero at any point.  
As $\sclup$ diminishes,
the two spin densities occupy the same coordinate
space, but on two very distinct length scales.
Even when $\sclup\to 0$, the up electrons
do not vanish, but are merely spread over
a very large volume.
The scaled density presumably then has vanishingly
small contribution to the correlation energy.
For finite systems, 
we can consider this limit as the effective removal 
of one spin density to infinitely far way.  
We will discuss what this means for extended systems later when we treat
the uniform gas.

Another interesting property of the spin scaling 
transformation is that a scaling of one spin 
density can always be written as a total density 
scaling plus an inverse spin scaling of the other spin; that is
\ben
E\xc[\n\ups,\n\dn]
=
E\xc[ \{\; \n\up,
\n_{\downarrow 1 / \sclup}
 \;\}_\sclup ]
\label{inv_ss_law}
\een 
where the parenthesis notation on the right indicates 
scaling the total density.  Thus, without loss of 
generality, we need only scale one spin density.  

To understand what happens when a single spin density 
is scaled, we first study exchange.  Because the spin 
up and down Kohn-Sham orbitals are independent, 
the exchange energy functional can be split into 
two parts, one for each spin \cite{PK98}.  
The scaling 
relationships for total \dft 
generalize for each term independently.  
For an up spin
scaling, we find 
\bea 
E\x [ \n\ups, \n\dn] &=&
\half E\x [ 2 \n\ups] 
+ \half E\x [ 2 \n\dn] \nonumber\\ 
&=& \frac{\sclup}{2} E\x [ 2\n\up] +
\half E\x [ 2 \n\dn] .
\label{ex_split}
\eea 
When
$\sclup\rightarrow 0$, we are left with only the
down contribution to exchange.
Separate spin scaling allows us to
extract the contribution from each spin density
separately, e.g., 
$ dE\x[\n\ups,\n\dn]/d \sclup$
at $\sclup=1$ is the contribution to the exchange
energy from the up density.
A plot of $E\x [ \n\ups, \n\dn]$ versus $\sclup$ between
0 and 1 yields a straight line and is twice as negative
at 1 as at 0.

Separate spin-scaling of the correlation energy is more
complicated.
Unlike $E\x[\n\up,\n\dn]$, 
$E\c[\n\up,\n\dn]$ cannot
trivially be split into up and down parts.  
The Levy method of scaling the exact ground-state 
wave-function does not yield an inequality such as Eq. (\ref{ec_scaling})
because the spin-scaled wave-function is not
a ground-state of another Coulomb-interacting Hamiltonian.
Nor does it yield an equality as in the spin-scaled exchange
case, Eq. (\ref{ex_split}), because the many-body wave-function is
not simply the product of two single spin wave-functions.
In both cases, the two spins are coupled by a term
$ 1 / |\br-\sclup\br'| $.

To obtain an exact spin scaling relationship for
$E\xc$, we take a different route.  Consider a
change in the energy due a small change in the
up-spin
density:
\ben 
\delta
E\xc= 
E\xc [\n\up +\delta\n\up , \n\dn ] 
- E\xc [\n\up, \n\dn ]. 
\label{delexc1}
\een 
Use 
$\pot\xcup (\br) =
\delta E\xc / \delta \n\up (\br)$ 
to rewrite
$\delta E\xc$ as 
\ben 
\delta E\xc  
= \int\d3r
\delta\n\up (\br)
\;\pot\xcup [\n\up,\n\dn] (\br) .
\label{delexc2}
\een
to first order in $\delta n\up$.  
Now, consider this change as coming from the following 
scaling of the density,  
$\n\ups (\br) = \sclup^3\n\up
(\sclup\br)$, where $\sclup$ is arbitrarily close to one.  
The change in the density is related to the derivative 
of this scaled density: 
\bea
\frac{d\n\ups (\br)}{d\sclup}|_{\sclup=1} = 3
\n\up ( \br) +
\br\cdot\nabla\n\up (\br) .
\label{derivn}
\eea
Use Eqs. (\ref{delexc2}) and (\ref{derivn}), and integrate
by parts to find
\ben 
\frac{d E\xc[\n\ups,\n\dn]}{d\sclup} |_{\sclup=1}=
-\int\d3r
\n\up(\br)\; \br\cdot \nabla
\pot\xcup[\n\up,\n\dn] (\br). 
\label{spin_virial_returns}
\een 
Eq. (\ref{spin_virial_returns}) is an exact result showing 
how $ dE\xc / d\sclup|_{\sclup=1} $
can be extracted from the spin densities and potentials.  
For an initially unpolarized system, $\n\up=\n\dn=\n/2$,
and $v\xcup=v\xcdn=v\xc$.  Thus the right-hand-side of
Eq. (\ref{spin_virial_returns}) becomes half the usual virial
of the exchange-correlation potential.
This virial  is equal
to $dE\xc[\n_{\sclup}]/d\sclup|_{\sclup=1}=E\xc+T\c$.
$T\c$ is the kinetic contribution to the
correlation energy\cite{LP85}.
Thus, for spin-unpolarized systems,
\ben 
\frac{d E\c[\n\ups,\n\dn]}{d\sclup} |_{\sclup=1}=
\half \left( E\c + T\c \right).
\label{dEcfromEcTc}
\een
For initially polarized systems, there is no simple relation
between the two types of scaling.

To generalize Eq. (\ref{spin_virial_returns}) to finite
scalings, simply replace $n\up$ on both sides by $\n\ups$,
yielding:
\ben 
\frac{d E\xc[\n\ups,\n\dn]}{d\sclup}=
-\frac{1}{\sclup}\int\d3r
\n\ups(r)\; \br\cdot \nabla
\pot\xcup[\n\ups,\n\dn] (\br). 
\label{dEcalph}
\een
We can then write the original spin-density functional
as a scaling integral over
this derivative:
\ben
E\xc[\n\up,\n\dn]=
\lim_{\sclup\rightarrow 0} E\c[\n\ups ,\n\dn ]+
\int_0^1 d\sclup \frac{d E\xc[\n\ups,\n\dn]}{d\sclup}.
\label{exc_spinint}
\een
This is a new expression for the exchange-correlation
energy as an integral over separately spin-scaled densities
where the spin-scaled density is scaled to the low-density limit.
With some physically reasonable assumptions, we expect
\ben
\lim_{\sclup\rightarrow 0} E\c[\n\ups ,\n\dn ] =
E\c[0,\n\dn ].
\label{ec_limit}
\een 
For example, if the anti-parallel correlation hole
vanishes as rapidly with scale factor as the parallel-spin correlation
hole of the scaled density, this result would be true.
Numerical results indicate that 
this is the case for the approximate functionals used in this
paper.  Nevertheless, we have not proven Eq. (\ref{ec_limit}) here.

A
symmetric formula can be written down by
scaling the up and down spins separately and
averaging:  
\bea 
E\xc&&[\n\up, \n\dn] 
= 
\half\lim_{\sclup\rightarrow 0} 
\left(
E\xc [\n\ups,\n\dn]
 + 
E\xc [\n\up, \n_{\downarrow \sclup }]
\right)
\nonumber\\ 
&&+\half
\int_0^1 d\sclup \int\d3r \n\ups(\br)
\br\cdot \nabla \pot\xcup [\n\ups,\n\dn] (\br) \nonumber\\
&&+\half 
\int_0^1 d\scldn \int\d3r
\n\dns(\br) \br\cdot \nabla
\pot\xcdn[\n\up,\n\dns] (\br) .
\label{excsym}
\eea 
This result is the spin density functional generalization of
spin-decomposition, coordinate scaling, and
the virial theorem.   Each of these ideas yields separate results
for pure exchange or uniform coordinate scaling, but
all are combined here.  
Notice that the potentials depend on both spins, 
one scaled and the other unscaled.  This reflects 
the difficulty in separating up and down spin correlations.  

The proof of Eq. (\ref{excsym}) is true for 
exchange-correlation, but  
in taking the weakly-correlated limit,
the result also holds true for exchange.  
In the exchange case, Eq. (\ref{excsym}) reduces to
Eq. (\ref{ex_split}) with equal contributions from the
limit terms and the virial contributions.  To obtain this result,
recall how $E\x$ scales, Eq. (\ref{scaling_ex}).
Since the energy contribution from each spin is separate 
and since the scaling law is linear, the limits in the first two terms 
of Eq. (\ref{excsym}) are doable without any extra physical assumptions.  
The virial terms are a bit more difficult to handle 
as the exchange potentials change under scaling.  
In the end, the first two terms contribute half the exchange 
energy while the virial terms
contribute the other half.  

\section{Uniform gas}
\label{s:unif}
%
\begin{figure}
\unitlength1cm
\begin{picture}(12,6.5) 
\put(0.2,0.0) 
{\psfig{figure=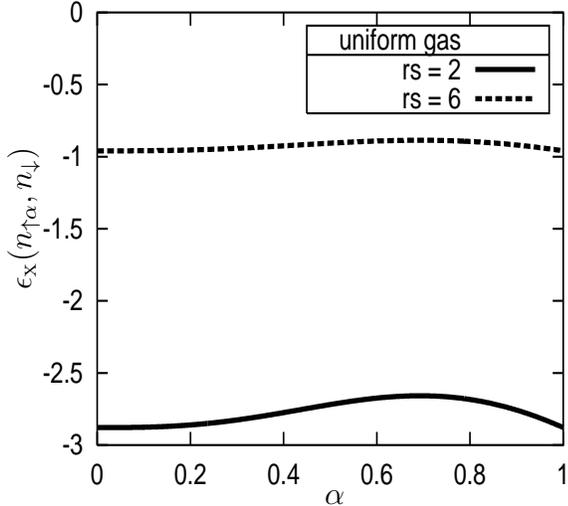,width=8cm,height=7cm}}
\setbox6=\hbox{\large $\epsilon\x (\n\ups,\n\dn)$}
\put(0.5,3.8){\makebox(0,0){\rotl 6}}
\put(4.5,0.0){\large $\sclup$}
\end{picture}
\caption{Spin scaling of a uniform gas: 
exchange energy per particle,
$\epsilon\x (\n\ups,\n\dn)$,
at $r_s=2$ (dotted line) and $6$
(solid line).}
\label{f:ssunifex} 
\end{figure}
\begin{figure}
\unitlength1cm
\begin{picture}(12,6.5) 
\put(0.2,0.0) 
{\psfig{figure=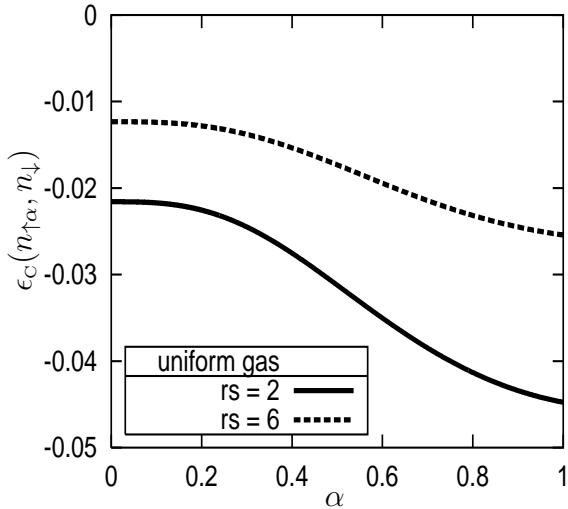,width=8cm,height=7cm}}
\setbox6=\hbox{\large $\epsilon\c (\n\ups,\n\dn)$}
\put(0.5,3.8){\makebox(0,0){\rotl 6}}
\put(4.5,0.0){\large $\sclup$}
\end{picture}
\caption{Spin scaling of a uniform gas: 
correlation energy per particle,
$\epsilon\c (\n\ups,\n\dn)$,
at $r_s=2$ (dotted line) and $6$
(solid line).}
\label{f:ssunif} 
\end{figure}

To illustrate the effect of spin scaling, 
we examine the uniform electron gas, a system
for which we have essentially exact results.
Great care must be taken to define quantities
during separate spin-scaling of extended systems.
Begin with a spin-unpolarized uniform electron gas of density 
$\n$ and
Wigner-Seitz radius
$r_s = ({3}/{4\pi\n})^\frac{1}{3}$.
When one spin
density is scaled, the system becomes spin-polarized,   
and relative spin-polarization is measured by
\ben
\zeta= \frac{
\n\up -\n\dn }
{\n\up +\n\dn }.
\label{zeta}
\een
We assume that for a spin-polarized uniform system,
the exchange-correlation energy per electron,
$\epsilon\unif\xc(r_s,\zeta)$, is known exactly.
We use the correlation energy parameterization 
of Perdew and Wang \cite{PW92} to make our figures.

To perform separate spin scaling of this system,
we focus on a region deep in the interior of any
finite but large sample.  A simple example is a jellium sphere
of radius $R >> r_s$. The correlation energy density 
deep in the interior will tend to that
of the truly translationally 
invariant uniform gas as $R\to\infty$.
At $\sclup=1$, we have an unpolarized system 
with $\n\up=\n\dn=\n/2$. 
The up-spin scaling, $\n\up = \sclup^3 n / 2$, changes
both the total density and the spin-polarization.  Deep in
the interior
\ben
r_s (\sclup)=r_s \left(\frac{2}{1+\sclup^3}\right)^{1/3}
\label{rs2}
\een
where $r_s$ is the Seitz radius of the original unpolarized gas,
and
\ben
\zeta(\sclup) = \frac{\sclup^3-1}
{\sclup^3 +1}.
\label{zeta2}
\een
The energy density  here
is then
\ben
e\xc(\sclup)=e\xc\unif(\n\ups,\n\dn)=e\xc\unif(r_s(\sclup),\zeta(\sclup)),
\label{excdeep}
\een
and the energy per particle is
\ben
\epsilon\xc(\sclup)= e\xc(\sclup)/\n(\sclup)
\een
where $\n(\sclup)$ is the interior density.

To illustrate the effects of this spin scaling,
consider the
simple exchange case.  
Deep in the interior, we have a uniform gas of density
$\n\ups$ and $\n\dn$, and the energy densities of these two
are given by
Eq. (\ref{ex_split}), since the integrals provide simple volume
factors.
The Slater factor of $\n^{4/3}$ in the exchange density
of the uniform gas produces a factor of $(1+\sclup^4)$.
When transforming to the energy per electron,
there is another factor of $(1+\sclup^3)$ due to the
density out front.  Thus the exchange energy per electron is
\ben
\epsilon\x (\sclup)
=
 \left(\frac{1+\sclup^4}
{1+\sclup^3} \right) \;
\epsilon\x^{unpol.} (\n)
\label{epsrel2}
\een
This variation is shown in Fig. \ref{f:ssunifex}.
This result may appear to disagree with Eq. (\ref{ex_split}),
but it is valid deep in the interior only.  To recover
the total exchange energy, one must include those electrons
in
a shell between $R$ and $R/\sclup$ with the full polarized 
uniform density $\sclup^3 n/2$.  
The exchange energy integral includes this contribution, and then
agrees with
Eq. (\ref{ex_split}).

Near $\sclup=1$, Eq. (\ref{epsrel2}) yields 
$(1+\sclup)\; \epsilon\x^{unpol.} / 2$, in agreement
with a naive application of Eq. (\ref{ex_split}).
This is because, in the construction of the energy
from the energy per electron,
the factor of the density accounts for changes in the
number electrons to first order.
So the derivative at $\sclup=1$ remains a good measure of
the contribution to the total exchange energy from one
spin density.
On the other hand, as $\sclup\to 0$, the exchange
energy per electron in the interior returns to that
of the original unpolarized case.  This reflects
the fact that exchange applies to each spin separately,
so that the exchange per electron of the down-spin density
is independent of the presence of the up-spin density.

Figure \ref{f:ssunif} shows the uniform 
electron gas correlation energy per particle
scaled from unpolarized ($\sclup=1$) to 
fully polarized limits ($\sclup=0$).  
Again, the curves become flat as $\sclup\to 0$,
because for small $\sclup$, there is very little
contribution from the up-spins.  Now, however,
there is a dramatic reduction from $\sclup=1$
to $\sclup=0$ because of the difference
in correlation between unpolarized and fully
polarized gases.  Note that the correlation
changes tend to cancel the exchange variations.


\section{Finite Systems}
\label{s:finite}


Next, we examine the behavior of finite systems
under separate-spin scaling.  We choose the He and Li
atoms, to demonstrate the effects on the simplest
non-trivial unpolarized and spin-polarized cases.
For each system, we solve the \KS
equations using a specific density functional
approximation.
The
resulting self-consistent densities are then spin
scaled and the approximate energies evaluated on the
scaled densities using that same functional.  
Since these are approximate functionals, neither the densities nor
the energies are exact.
We are unaware of any system, aside from the uniform gas, 
for which exact spin-scaled plots are easily obtainable.
For now, we must compare plots generated from
approximate functionals.  
Even the simple atomic calculations presented here were
rather demanding since, especially for very small spin-scaling
parameters, integrals containing densities
on two extremely distinct length scales are needed.


\begin{figure}
\unitlength1cm
\begin{picture}(12,6.5) 
\put(0.2,0.0) 
{\psfig{figure=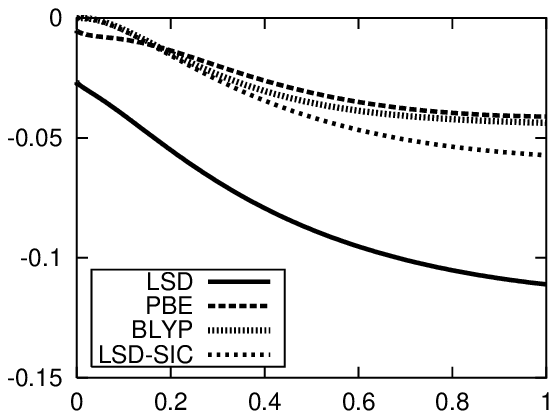,width=8cm,height=7cm}}
\setbox6=\hbox{\large $E\c[\n\ups,\n\dn]$}
\put(0.5,3.8){\makebox(0,0){\rotl 6}}
\put(4.5,0.0){\large $\sclup$}
\end{picture}
\caption{
Spin scaling of the He atom density using various approximate
functionals for $E\c$: \spielplus .} 
\label{f:ssecHe}
\end{figure}

The He atom (Fig. \ref{f:ssecHe}) is spin 
unpolarized at $\sclup=1$.  Scaling either spin 
density gives the same results.
The LSD curve gives far too much correlation 
and does not vanish as $\sclup\to 0$.
In the fully scaled limit, we expect, 
as we have argued in section \ref{s:theory}, that 
the correlation energy
should vanish. 
This is because the two electrons are now on
very different length scales and so should
not interact with each other.  The residual
value at $\sclup \rightarrow 0$ reflects the self-interaction
error in LSD for the remaining
(unscaled) one-electron density.
The PBE curve is on the right scale but also has
a residual self-interaction error as $\sclup\to 0$.
The BLYP functional \cite{B88,LYP88} is popular
in quantum chemistry and gets both limits correct.
However, the functional's lack of self-interaction error is 
because the correlation energy vanishes for {\em any}
fully polarized system.  This vanishing is incorrect for 
any atom other than H or He.
Finally, the LSD-SIC curve \cite{PZ81} is probably the most
accurate in shape (if not quantitatively) since this
functional handles the self-interaction error
appropriately.
We further observe that the curves appear quite different
from those of the uniform gas.  The atomic curves are much
flatter near $\sclup\rightarrow 1$ and have appreciable
slope near $\sclup \rightarrow 0$. This is because these
energies are integrated over the entire system, including
the contribution from the entire spin-scaled density,
whereas the energy densities in the uniform gas case were
only those in the interior.

\begin{table}
\begin{center}
\begin{tabular}{|lcccc|} 
\hline
Approx.	& $E\x$  &  $E\c$ & $E\c[\n\up,0]$ & $dE\c/d\sclup$ \\ \hline
LSD 	& -0.862 & -0.111 & -0.027         & -0.022         \\
PBE 	& -1.005 & -0.041 & -0.005         & -0.002         \\
SIC 	& -1.031 & -0.058 &  0.000         & -0.011         \\
BLYP	& -1.018 & -0.044 &  0.000         & -0.005         \\
exact	& -1.026 & -0.042 &  0.000         & -0.003         \\ \hline
\end{tabular}
\caption{\label{t:heectable}
He atom energies, both exactly and within several approximations.
All energies in Hartrees; all functionals evaluated on self-consistent
densities.}
\end{center}
\end{table}
Quantitative results are listed in Table \ref{t:heectable}.
The exact He values, 
including the derivative at $\sclup=1$, using 
Eq. (\ref{dEcfromEcTc}), 
were taken from Ref. \cite{UG94,FTB00}.
Note that PBE yields the
most accurate value for this derivative.  
The BLYP correlation energy 
is too flat as function of scale parameter.  BLYP produces
too small a value for $T\c$ leading to a lack
of cancellation  with $E\c$ and a subsequent  overestimate
of the derivative at $\sclup=1$.  LSD-SIC has
a similar problem.  The LSD value, while far too large,
is about 8\% of the LSD correlation energy, close to the
same fraction for PBE, and not far from exact.
However, the important point
here is that results from separate spin-scaling are a new
tool for examining the accuracy of the treatment of
spin-dependence in approximate spin-density functionals.

\begin{figure}
\unitlength1cm
\begin{picture}(12,6.5) 
\put(0.2,0.0) 
{\psfig{figure=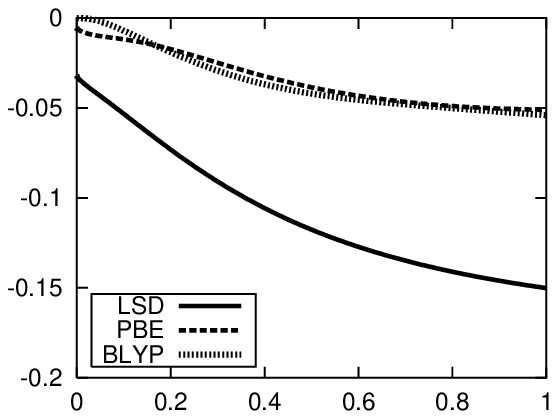,width=8cm,height=7cm}}
\setbox6=\hbox{\large $E\c[\n\ups,\n\dn]$}
\put(0.5,3.8){\makebox(0,0){\rotl 6}}
\put(4.5,0.0){\large $\sclup$}
\end{picture}
\caption{Up spin scaling 
of the Li atom density using various approximate
functionals for $E\c$: \spiel .}
\label{f:ssecliup} 
\end{figure}
\begin{figure}
\unitlength1cm
\begin{picture}(12,6.5) 
\put(0.2,0.0) 
{\psfig{figure=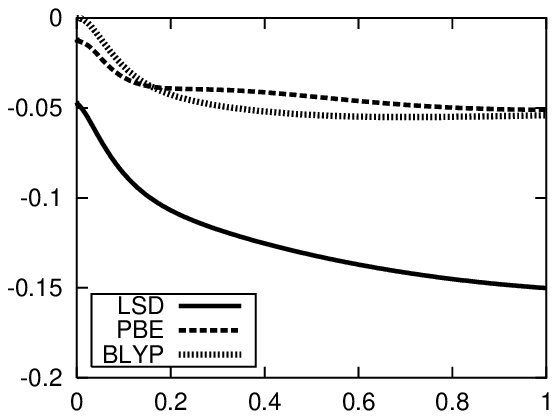,width=8cm,height=7cm}}
\setbox6=\hbox{\large $E\c[\n\up,\n\dns]$}
\put(0.5,3.8){\makebox(0,0){\rotl 6}}
\put(4.5,0.0){\large $\scldn$}
\end{picture}
\caption{Down spin scaling 
of the Li atom density using various approximate
functionals for $E\c$: \spiel .}
\label{f:sseclidn} 
\end{figure}

The Li atom (Figs.\ref{f:ssecliup} 
and \ref{f:sseclidn}) is the smallest non-trivial 
odd-electron atom.   We choose the up spin density
to have occupation \emph{1s2s}.  As the up spin is scaled
away, as in Fig. \ref{f:ssecliup}, we  find a curve
very similar to that of He, Fig. \ref{f:ssecHe}.
The primary difference is the greater correlation
energy for $\sclup=1$.  

On the other hand, scaling away the down-density gives
a very different picture, Fig. \ref{f:sseclidn}.
The most dramatic changes in the correlation 
energy now occur at small $\sclup$.  Near
$\sclup\rightarrow 1$, the system energy is quite insensitive to
spin-scaling, especially in GGA.  This is exactly opposite 
to what we have seen for the uniform gas.  It is an open 
question whether this would be observed with the exact functional.
For up spin scaling, we expect the correlation energy to vanish as
$\sclup\rightarrow 0$.
But for down spin scaling one expects a finite correlation energy
in the limit $\scldn \rightarrow 0$. The two spin-up
electrons remain and are still correlated.  In this case, 
the BLYP functional errs noticeably since it predicts no correlation 
energy for the remaining two electrons.

\begin{table}
\begin{center}
\begin{tabular}{|lcccccc|} 
\hline
Approx.	& $E\x$  &  $E\c$  & $E\c[\n\up,0]$ & $dE\c/d\sclup$ & $E\c[0,\n\dn]$ & $dE\c/d\scldn$  \\ \hline
LSD     & -1.514 & -0.150 & -0.047        & -0.037        & -0.032        & -0.019         \\  
PBE     & -1.751 & -0.051 & -0.012        & -0.004        & -0.005        & -0.001         \\ 
BLYP    & -1.771 & -0.054 & -0.054        & -0.020        &  0.000        & -0.005         \\ 
exact   & -1.781 & -0.046 & -             &  -            &  0.000        &  -             \\ \hline
\end{tabular}
\caption{\label{t:liectable}
Li atom energies, both exactly and within several approximations.
All energies in Hartrees, all functionals evaluated on self-consistent
densities.  }
\end{center}
\end{table}
Quantitative results for Li are given in Table \ref{t:liectable}.
The exact result for $E\x$ is the $E\x$ of a self consistent OEP calculation.
Using the highly accurate energy prediction from \cite{DHCUF91}, 
we deduce the exact $E\c = E_{T} - E_{T,OEP}$.  
The other exact results are not extractable from the literature here,
but could be calculated from known exact potentials
and densities \cite{MZ95}.  
Even in this simple case, an SIC calculation
is difficult.  For the up-spin density,
one would need to find the \emph{1s} and \emph{2s} 
orbitals for each value of $\sclup$ 
that yield the spin-scaled densities.

\section{Spin adiabatic connection}
\label{s:adia}

Here, we define an analog of the
adiabatic connection within
the spin-scaling formalism. Traditionally, we think
of $\lambda$ as a parameter in the Hamiltonian, but
this way of thinking becomes prohibitively complicated in
spin density functional theory.  
We would have to define three coupling
constants:  $\lambda\up$, $\lambda\dn$, and
$\lambda_{\uparrow\dnarrow}$.  
Even if we did that, it would be non-trivial 
to relate changes in these coupling constants to changes 
in the electron density.
Instead, we {\em define} a relationship 
between spin-scaling and a spin dependent \emph{coupling parameter}.  
For total density scaling, the
relationship between scaling and evaluating a
functional at a different coupling constant is \cite{LP85,Wb97} 
\ben
E\xc^\lambda[\n] =\lambda^2 E\xc[\n_{1/\lambda}].
\label{ccandscale}
\een
The adiabatic connection formula is
\ben
E\xc= \int_0^1d\lambda\ \frac{d E\xc\l}{d\lambda} 
= \int_0^1d\lambda\ U\xc(\lambda) .
\label{excacf}
\een
By virtue of the Hellmann-Feynman
theorem, $U\xc(\lambda)$ can be identified as the 
potential contribution to exchange-correlation at
coupling constant $\lambda$.  
The integrand $U\xc(\lambda)$ can be plotted
both exactly and within density functional approximations,
and its behavior lends insight 
into deficiencies of functionals \cite{BPEb97}.
For separate spin scaling, we apply the same ideas
but now to 
\ben
\Delta E\xc[\n\up,\n\dn]=E\xc[\n\up,\n\dn]-
E\xc[0,\n\dn],
\label{excdiff}
\een
the exchange-correlation energy difference between
the physical system and the system with one spin density removed 
while keeping the remaining spin-density fixed.
For polarized systems, this quantity depends on which
spin density is removed. 
We define
\ben
\Delta E\xc^{\lamup} = \lamup^2 
\Delta E\xc [\n_{1/\lamup},\n\dn]
\label{delexc_def}
\een 
and 
\ben 
\Delta U\xc(\lamup)  =
{d\Delta E\xc^{\lamup} }/
{d \lamup },
\label{uxc_spin}
\een
so that
\ben
\Delta E\xc = \int_0^1 d\lamup\ \Delta U\xc(\lamup).
\label{exclam}
\een
This produces a spin-dependent decomposition of the
exchange-correlation energy, related to separate
spin-scaling rather than total density scaling,
with the integral now including the high-density
limit.
As $\lamup\to 0$, exchange dominates, and
$U\xclup \to U\xlup$ which is just $E\x[2\n\up]/2$ 
according to the simple results for exchange in Sec. \ref{s:intro}.
Furthermore, in the absence of correlation, $U\xclup$ is
independent of $\lamup$.  This is not true if one uses
a naive generalization of Eq. (\ref{ccandscale}).

This spin adiabatic connection
formula should prove useful for the improvement of present-day
functionals in the same way that the 
adiabatic connection formula has been useful for
improving total density functionals.
For example, it might be possible to perform G\"orling-Levy perturbation
theory \cite{GL94}
in this parameter ($\lamup$) or to extract a correlation contribution
to kinetic energy \cite{Sc96}.
\begin{figure}
\unitlength1cm
\begin{picture}(12,6.5) 
\put(0.2,0.0) 
{\psfig{figure=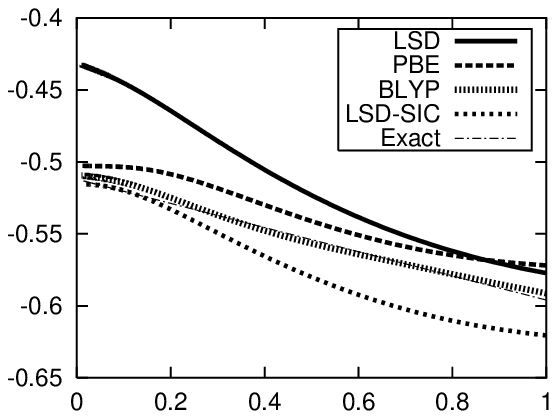,width=8cm,height=7cm}}
\setbox6=\hbox{\large $ \Delta U\xclup $}
\put(0.5,3.8){\makebox(0,0){\rotl 6}}
\put(4.5,0.0){\large $\lamup$}
\end{picture}
\caption{Single spin adiabatic connection for He atom: 
\spielplus , Exact (fancy dashes).}
\label{f:he_adiacon} 
\end{figure}
\begin{table}
\begin{center}
\begin{tabular}{|lccccc|} 
\hline
                 &  LSD  &  PBE    & SIC    & BLYP   &  exact \\  \hline 
$\Delta U\xc(0)$ & -0.43 & -0.50   & -0.53  & -0.51  & -0.51  \\  
$\Delta U\xc(1)$ & -0.58 & -0.57   & -0.62  & -0.59  & -0.60  \\ 
$\Delta E\xc$    & -0.54 & -0.54   & -0.57  & -0.55  & -0.56  \\ \hline
\end{tabular}
\caption{\label{t:headia}  Spin adiabatic connection  $\Delta U\xclup$
for He atom, both exactly and in several approximations.}
\end{center}
\end{table}
We show the spin adiabatic connection for 
the He atom in Fig. \ref{f:he_adiacon}.  In generating
each adiabatic connection plot, we now take the scaled spin density to
the high density limit.
The area under each curve  is precisely $\Delta E\xc$ for a particular
approximation.  To get $E\xc[\n\up,\n\dn]$,  we must add  
the  contribution  from   the  unscaled  spin,
$E\xc[0,\n\dn]$.   The spin  adiabatic  connection curve looks quite similar  
to the regular adiabatic connection curve:
for the He atom, $\Delta U\xclup$ becomes more negative
with $\lambda$ everywhere and is close to linear.
This suggests that the spin-correlation effects are weak
for this system, just as the correlation effects are.

To better understand how popular approximations perform,
we would like to compare with the exact curve.  In principle, this
requires a sophisticated wavefunction calculation designed to reproduce
the spin-scaled densities at every point in the adiabatic connection
curve.  Here, we use a simple interpolation that should be highly accurate.
Analytic formulae
give exact limits for $\Delta U\xclup$.  
At the small $\lamup$ limit, exchange  dominates, 
and  we are  left  with the
exchange contribution from the scaled spin to the total energy: 
\ben
\Delta U\xc (\lamup=0)=
\half E\x[\n]
\label{DUxcat0}
\een
At the other end,
\bea
\Delta U\xc(\lamup = 1) 
= 2  E\xc[\n\up,\n\dn] -2 E\xc[0,\n\dn]
\nonumber\\  
-{d E\xc[\n\ups,\n\dn]}/{d\sclup} |_{\sclup=1}
\label{DUxcat1}
\eea
For a spin-unpolarized two electron system like the
He atom, this becomes
\bea
\Delta U\xc(\lamup=1)=E\x/2+2 E\c - (E\c+T\c)/2 \nonumber \\
~~~~~(\mbox{2 electrons}, \mbox{unpol.})
\eea
For He at $\lamup = 1$, $\Delta U\xclup = -0.60$.  
To approximate the exact curve, we use a (1,1) Pad\'e
approximant.
The values
$\Delta U\xc(0)$, $\Delta U\xc(1)$, and $\Delta E\xc$ 
fix the three unknown parameters.  This pade turns out to be nearly 
a straight line.  

Table \ref{t:headia}  shows the exact  limits and the limits  given by
several  popular functionals.  
BLYP reproduces both limits most accurately 
and is mostly linear. This should  come as no surprise
as  BLYP yields good energies and accounts for  He's self-interaction 
error (if a bit serendipitously).  However,  we do  not
expect  such good  results from  BLYP  when using it on Li.  For Li, 
as we have see in section \ref{s:finite}, BLYP predicts 
no correlation energy when only one electron is scaled away.  BLYP will 
fail noticeably and uncontrollably in this case.
The LSD functional 
dramatically underestimates the single spin exchange energy 
and, therefore, gets the small $\lamup$ limit quite wrong.  
This reflects the usual error for LDA exchange.
But notice how well LSD performs performs at $\lamup=1$.
The value here is only a 3\% overestimate of the exact value,
much better than the 9\% overestimate for the exchange-correlation energy.
Furthermore, the LSD derivative as $\lamup\to 1$ is almost exact.
PBE and LDA-SIC  are qualitatively  similar, the greater error in
LSD-SIC being due to the errors in LSD.  Both show a flattening
of the curve as $\lamup\to 0$, much more than BLYP.  
Our \emph{exact} 
curve is too crudely constructed to indicate which behavior is more accurate.

Ideally, we would compare approximations to the exact adiabatic plot 
for this and other systems such as the Li atom.  The plots are not easy to generate. 
But even so, analysis of the exact limits
is sufficient to garner a deeper understanding of how functionals 
treat and mistreat spin densities.
 

\section{Conclusions}

Both scaling and the adiabatic decomposition formula have 
proven extremely useful in studying and constructing total
density functionals.
We have suggested the possibility of scaling spin densities
separately, derived a new virial theorem, given new exact
results for the He atom,
and pointed out the difficulties of deducing exact theorems
from this decomposition.
While exact calculations are difficult to perform and
and exact results appear difficult to prove within this 
approach, any results would be very useful and likely to improve
spin density functional theory's treatment of magnetic properties.

We close with a significant challenge to developing
separate spin scaling.   In the total density scaling of
Eq. (\ref{scaling_law}), the
density is both squeezed (or spread)
and is also translated.   
The squeezing is
independent of the choice of origin, but the
translation is not. This
origin-dependence should not affect 
the exchange-correlation energy
because 
space is translationally invariant.
However, when
an individual spin density is
scaled, the remaining spin density remains fixed in space.
This means the resulting density depends on
the choice of origin for the separate spin-scaling.
So while $E\c[\n\ups, \n\dn]$ is a spin-density functional of
$\n\ups$ and $\n\dn$, it is {\em not} a pure spin-density functional
of the original spin-densities because of this origin dependence.
Most likely, a method of transforming away this origin
dependence, as found for virial energy 
densities in Ref. \cite{BCL98}, will be needed to make this spin scaling 
technique more physical
and useful.  
For atoms, we made the obvious choice of origin at the 
center of the nucleus.  Origin dependence will become
acute in applications to molecules and even worse for solids.
On the other hand, the non-uniform coordinate scaling of
G\"orling and Levy \cite{GL92} suffers from the same
difficulties for non-spherical densities but has
still produced useful limits for approximate
density functionals \cite{BPL95}.

However, it is important to stress that 
the spin virial relationship is unaffected by this challenge.
For $\alpha$ arbitrarily close to 1,
the spin-scaled energies are independent of the choice of origin, 
and these 
difficulties are irrelevant.  The spin virial relationship 
is an 
exact constraint and
give us a useful measure of how the correlation energy is affected by 
small changes in the spin densities.  It also leads to a natural 
decomposition of energy changes due to separate spin densities.  
It should be useful in determining whether calculations are 
self-consistent 
for each spin density separately.  This might be useful for example 
in systems where small differences between spin densities are 
important to calculate properly.


\section{Acknowledgments}

We thank Federico Zahariev for useful discussion, and
Eberhard Engel for use of his atomic density functional
code.
This work was supported by the National Science Foundation 
under grant number CHE-9875091. 
R. Magyar and T. Whittingham 
were funded by GAANN fellowships.


\end{document}